# Solid Phase Recrystallization in Arsenic Ion-Implanted Silicon-On-Insulator by Microsecond UV Laser Annealing

Toshiyuki Tabata, *Member*, *IEEE*, Fabien Rozé, Pablo Acosta Alba, Sebastien Halty, Pierre-Edouard Raynal, Imen Karmous, Sébastien Kerdilès, and Fulvio Mazzamuto

*Abstract*—UV laser annealing (UV-LA) enables surface-localized high-temperature thermal processing to form abrupt junctions in emerging monolithically stacked devices, where the applicable thermal budget is restricted. In this work, UV-LA is performed to regrow a silicon-on-insulator wafer partially amorphized by arsenic ion implantation as well as to activate the dopants. In a microsecond scale (~$10^{-6}$ s to ~$10^{-5}$ s) UV-LA process, monocrystalline solid phase recrystallization and dopant activation without junction deepening are evidenced, thus opening various applications in low thermal budget integration flows. However, some concerns remain. First, the surface morphology is degraded after the regrowth, possibly because of the non-perfect uniformity of the used laser beam and/or the formation of defects near the surface involving the excess dopants. Second, many of the dopants are inactive and seem to form deep levels in the Si band gap, suggesting a further optimization of the ion implantation condition to manage the initial crystal damage and the heating profile to better accommodate the dopants into the substitutional sites.

## I. Introduction

Nowadays, to further explore alternative scaling paths, monolithically stacked devices are emerging [1-8]. However, vertical stacking of multiple functional layers brings severe limitation of the thermal budget applicable to top-tier devices (e.g., 500 °C for 2 h [3,6]), because bottom-tier ones must preserve their functions and performances during subsequent thermal processing steps. One of the most critical challenges is the formation of junctions on the top layers (e.g., source and drain [9] and their extension [8] for metal–oxide–semiconductor field-effect transistors (MOSFET), back-surface passivation for backside-illuminated complementary MOS imager sensors (BSI-CIS) [10,11]). If conventional annealing processes such as high temperature furnace and rapid thermal processing (RTP) are not compatible, is generally used a lower temperature process called solid phase recrystallization (SPR), which can be performed at 500 to 600 °C in ion-implanted silicon (Si) [12,13]. However, such low-temperature SPR undergoes significant reduction of the recrystallization rate [14]. To compensate it, the annealing time must be extended, but it may result in deactivation of the dopants. Ion implantation introduces a high concentration of interstitials (i.e., point defects) underneath the amorphized layer, and a thermal processing transforms them into dislocation loops (i.e., extended defects) [15]. These so-called end-of-range (EOR) defects newly release Si interstitials towards the neighboring surface and interface to reduce the free energy of system [16,17]. Those Si interstitials may interact with substitutional dopants and lead to their deactivation [17].

UV laser annealing (UV-LA) can be advantageous for SPR because it allows to reach a high temperature while conserving the functionality of surrounding devices thanks to its short timescale and shallow irradiation absorption. In our previous work, a UV-LA SPR process using a nanosecond pulsed laser (its effective dwell time was ~$10^{-7}$ s) has been demonstrated on a 22-nm-thick silicon-on-insulator (SOI) substrate partially amorphized by ion implantation [18,19]. Then, a maximum crystallization rate of 1.8 nm per single pulse irradiation was obtained, and it was necessary to deposit thermally independent multiple laser pulses so that the entire amorphized layer was recrystallized. Considering a typical amorphization thickness in fully depleted SOI devices (e.g., 5 to 7 nm for the extension [8], 13 to 20 nm for the source and drain [1,17]), this crystallization rate is acceptable. However, other devices such as power IC may require recrystallization of a much thicker layer (e.g., 150 nm for a SOI-based lateral double-diffused MOSFET [20]), and a more effective control of recrystallization, particularly extending the dwell time of laser processing, may be mandatory to guarantee a reasonable productivity for high volume manufacturing.

In this work, we present a UV-LA SPR process performed with a microsecond-scale dwell time (i.e., ~$10^{-6}$ s to ~$10^{-5}$ s) on a relatively thick SOI structure amorphized by arsenic (As) ion implantation for roughly a half of its thickness. Sheet resistance, surface morphology, atomic diffusion, As activation, and SOI microstructure after UV-LA are discussed, evidencing a SPR completion among applied process conditions. The work initially presented in the 20th International Workshop on Junction Technology (IWJT2021) [21] has been re-edited, newly involving (i) dwell time as a parameter representing the UV-LA process, (ii) time-temperature profile simulation, and (iii) additional experimental data.

## II. EXPERIMENTAL PROCEDURE

A 70-nm-thick SOI wafer having a 145-nm-thick buried oxide (BOX) was used as starting material. The wafer had a (100) surface. Arsenic ion implantation was performed at room temperature (RT) at 19 keV with a total dose of $4 \times 10^{15}$ cm$^{-2}$. A 37-nm-thick amorphization was confirmed by cross-sectional transmission electron microscopy (TEM). The wafer was then submitted to UV-LA at RT in air, varying both laser fluence and dwell time to control the heat generated in the SOI structure. The laser wavelength was between 300 and 400 nm. Sheet resistance ($R_{sq}$) was measured by a standard four-point probe. Surface morphology was observed by atomic force microscopy (AFM). Arsenic and oxygen (O) diffusion within the SOI layer were measured by secondary ion mass spectroscopy (SIMS). Crystallinity after UV-LA was discussed by cross-sectional TEM. Active shallow donor and inactive defect-related deep donor/acceptor concentrations were estimated by an analytical model developed for electrochemical capacitance voltage profiling (ECVP) [22].

To extract the time-temperature profile in the ion-implanted SOI structure, the UV-LA process was simulated by means of a self-consistent time-harmonic solution of the Maxwell equations, whose self-consistency derives from dependence of an optical constant on temperatures and phases (i.e., amorphous, crystal, and liquid). More details of our simulation are explained elsewhere [23]. The dwell time is defined at the full-width half-maximum of the profile extracted for the amorphized SOI layer. It should be noted that our simulation model does not consider yet the progress of the SPR front, which changes the total reflectivity of the stack (i.e., the absorption efficiency of the irradiated laser light). For instance, the theoretically calculated reflectance is 0.501 for the as-implanted stack, whereas 0.576 for the fully recrystallized one. However, a rough estimation of the applied UV-LA heating dynamics may be still obtained.

## III. RESULTS AND DISCUSSION

First, based on the $R_{sq}$ value measured in the annealed SOI layer, we identified a SPR condition. As shown in Fig. 1, at a given laser fluence, increasing the dwell time led to a continuous decrease of the $R_{sq}$ value. Interestingly, the $R_{sq}$ value then became constant at around $10^2$ ohm/sq. Here, we suppose that the ion-implanted SOI layer is fully crystallized by SPR, and that the $R_{sq}$ value does not further decrease until the dopants start to diffuse due to melting of the SOI layer. In the following discussion, the sample annealed for 20 μs will be focused.

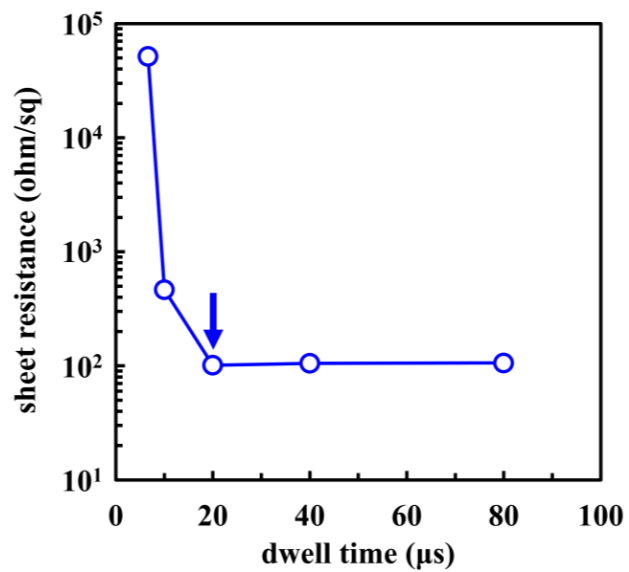

Fig. 1. Sheet resistance ($R_{sq}$) plots at a given laser fluence as a function of the dwell time.

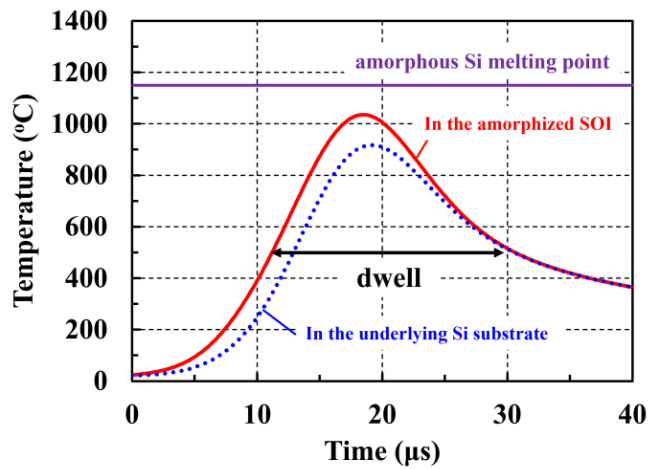

Fig. 2. Time-temperature profile obtained by our simulation for the 20 μs UV-LA process. The dwell time is defined at the full-width half-maximum of the profile extracted for the amorphized SOI layer. The amorphous Si melting point is also shown.

At this UV-LA condition, the time-temperature profile in the amorphized SOI part was simulated. As shown in Fig. 2, the maximum temperature reached during the process (~1030 °C) does not exceed the melting point of amorphous Si (1420 K [24]). Simply dividing the amorphized SOI layer thickness (37 nm) by the applied dwell time (20 μs), the overall SPR rate can be approximately estimated to be in a range of ~$10^{-3}$ m/s. This value is higher for almost two orders of magnitude than the one reported in the literature for the temperature close to ~1030 °C (~$10^{-5}$ m/s) [25]. It is indeed known that the doping enhances the SPR rate [26]. A thermal isolation of more than 100 °C in terms of the maximum temperature is also seen between the amorphized Si part and the underlying crystalline Si substrate.

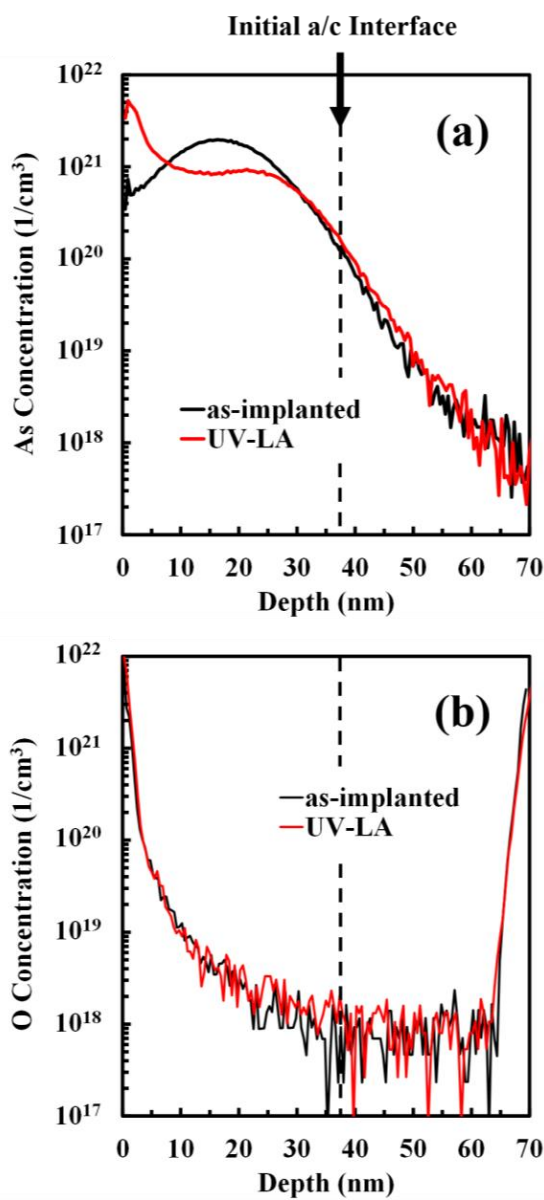

Fig. 3. SIMS profiles of (a) As and (b) O atoms before and after the 20 μs UV-LA. The initial amorphous/crystal (a/c) interface in the SOI layer is also indicated by an arrow and a dotted line.

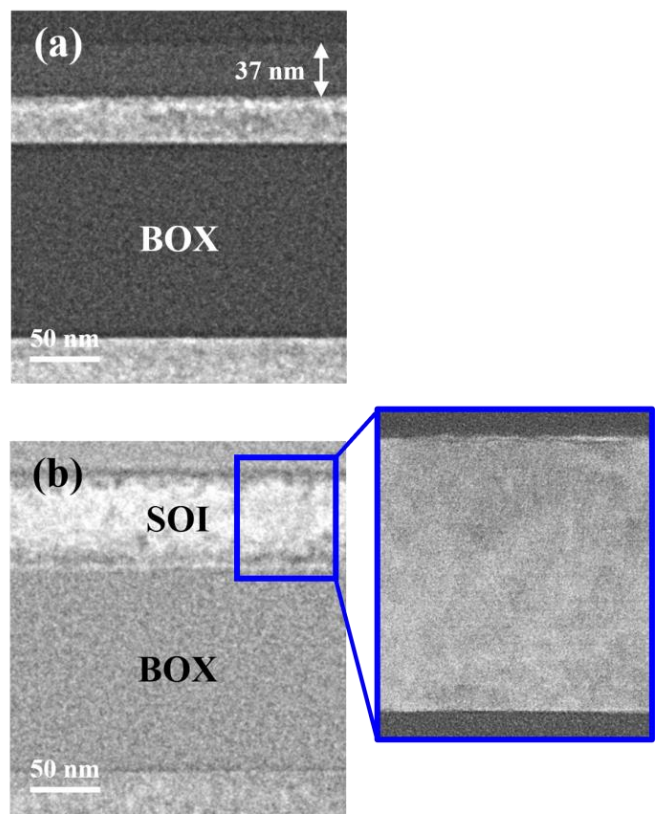

Fig. 4. Dark-field cross-sectional TEM images taken (a) before (i.e., as-implanted) and (b) after the 20 μs UV-LA process.

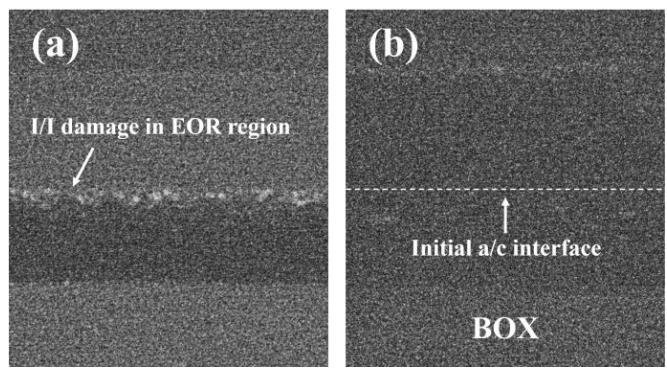

Fig. 5. Weak-beam dark-field cross-sectional TEM images taken (a) before (i.e., as-implanted) and (b) after the 20 μs UV-LA process. The observation was performed with g/3g, g=004, aiming at visualizing point defects (ion implantation damage) and their coalesced defects [32].

Inserting the 145-nm-thick silicon dioxide ($SiO_2$) layer helps it, because its thermal conductivity (~$10^0$ W/m.K at 300 K [27]) is much lower than that of crystalline Si (~$10^2$ W/m.K at 300 K [27]). The degree of this thermal isolation should be tunable by the UV-LA parameters such as the wavelength of laser light as well as the frequency and duration of irradiation if the laser is pulsed.

Fig. 3 shows the As and O SIMS profiles before and after the 20 μs UV-LA. An important evidence of SPR is that there is no O diffusion at this condition (Fig. 3(b)). When melting of an ion-implanted Si happens, O incorporation is observed even in a nitrogen ambient for a much shorter dwell time (~$10^{-7}$ s) [28]. On the other hand, at this possible SPR condition, a clear As segregation towards the surface was observed in the initially amorphized SOI layer (Fig. 3(a)). This may be induced by a "snowplow effect" in front of the moving amorphous/crystal (a/c) interface during SPR [29]. This characteristic As redistribution can be beneficial for lowering contact resistivity of transistors [30,31].

Fig. 4 shows the dark-field cross-sectional TEM images taken before and after the 20 μs UV-LA. A partial amorphization of the SOI layer by the As ion implantation is clear and the amorphized thickness can be estimated to be about 37 nm (Fig. 4(a)). After the UV-LA process, the amorphized part was fully recrystallized into the single-crystalline state, showing no grain boundary in the presented cross-sectional TEM images. (Fig. 4(b)). This observation finally confirms a SPR completion at this UV-LA condition. Here, some contrast non-uniformity was also observed in the whole SOI layer after

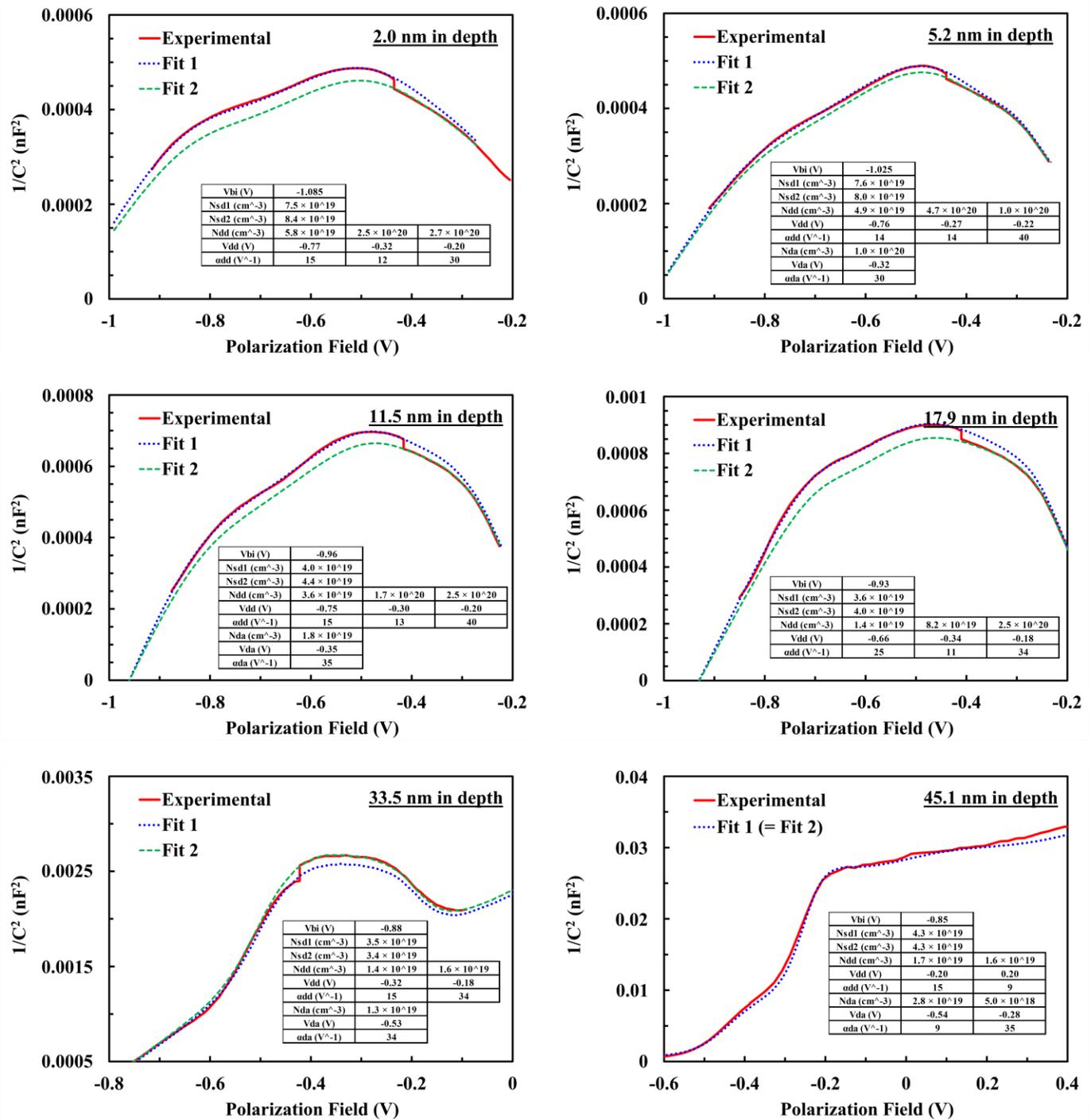

Fig. 6. Experimental $1/C^2$ ($C$ is the capacitance) versus the polarization field ($V$) curves (see the red solid lines) taken at different depths. The blue dotted and green dashed curves correspond to the fitting models (Fit 1 and Fit 2) with two different concentrations of shallow donors ($N_{sd1}$ and $N_{sd2}$, respectively). The other fitting parameters are $N_{dd}$ and $N_{da}$ for the concentration of deep donors and acceptors, $V_{dd}$ and $V_{da}$ for the polarization field at which the deep levels get ionized, then $α_{dd}$ and $α_{da}$ for the inverse of the potential variation inside the space charge zone [22].

its recrystallization, and this may imply presence of crystalline defects. It is possible that they were not visible in the presented TEM images because of their geometric relation to the applied analysis view [32] as well as of not high enough resolution. Then, the Si interstitials, whose source should be the ion implantation damage prior to UV-LA, probably do not contribute to formation of new crystalline defects during the SPR induced by the 20 µs UV-LA process. Considering the applied dwell time, the Si self-diffusion length calculated at the melting temperature of amorphous Si (1420 K [24]) in crystalline Si becomes negligible (~$10^{-3}$ nm [33]). Therefore, it is supposed that the Si interstitials cannot travel and introduce any self-coalescence. As shown in Fig. 5, the weak-beam dark-field images indicates that many point defects were left prior to the UV-LA treatment underneath the initial amorphous/crystal (a/c) interface (Fig. 5(a)), but no white contrast implying the presence of either of remaining point defects or coalesced defects was observed after the 20 µs UV-LA SPR in the whole

SOI layer (Fig. 5(b)). In the bottom-half of the SOI layer (i.e., non-amorphized), the contrast non-uniformity might be an artifact coming from our cross-sectional TEM lamella preparation, because it is present even near the SOI/BOX interface both before and after UV-LA SPR. On the other hand, in the upper-half of the SOI layer (i.e., amorphized then recrystallized by UV-LA SPR), the non-uniform contrast seems becoming darker compared to the bottom part in the magnified cross-sectional TEM image, implying another contribution in addition to the possible TEM artifact. Indeed, the fast placement of atoms at the moving a/c interface during UV-LA SPR might leave disorders in the regrown crystal. Furthermore, the solid solubility limit of As in Si (~3 × $10^{20}$ cm$^{-3}$ at 1100 °C [34]) and the As concentration depth profile obtained by SIMS (>3 × $10^{20}$ at 0 to 34 nm in depth (Fig. 3(a))) suggest formation of atomic clusters involving excess As atoms, and it might be enhanced towards the SOI surface along the As surface segregation. In fact, after the 20 μs UV-LA, the root-mean-square (RMS) value of surface roughness measured by AFM with a 30 μm × 30 μm scan was increased (~1.5 nm), compared to the as-implanted sample (~0.10 nm). This might be related to such atomic clusters preferably emerging near the surface. However, it should be noted that the laser beam used in this study was not perfectly optimized and still underwent a non-uniform shape. One may therefore speculate that a heat gradient is introduced in the SOI layer plane, resulting in a local variation of recrystallization rate. Anyhow, the possible formation of defects in the SOI layer during UV-LA SPR must be carefully investigated especially by means of high-resolution TEM to clearly capture their appearance.

It would be worth comparing our results with the others obtained by the LA treatment (regardless of UV or not) having different dwell times. In the case of a shorter dwell time (~$10^{-7}$ s) from our previous works [18, 19], no obvious defect was pointed out at the cross-sectional TEM level after a SPR completion in a phosphorus (P) (4 keV, 1 × $10^{15}$ cm$^{-2}$) [19] or As (9 keV, 1 × $10^{15}$ cm$^{-2}$) [18] doped 22-nm-thick SOI layer. As the calculated Si self-diffusion length is further reduced in this shorter dwell time (~$10^{-4}$ nm [33]), any self-coalescence of the Si interstitials (i.e., ion implantation damage) is not expected. In addition, since the as-implanted peak concentration of the dopants (~5 × $10^{20}$ cm$^{-3}$ for the applied P condition [35] and ~1 × $10^{21}$ cm$^{-3}$ for the As one [36]) was lower than that of this work (~2 × $10^{21}$ cm$^{-3}$), the possible atomic clustering would be greatly suppressed. A flat surface (~0.12 nm in RMS) was then maintained even after SPR [19]. Either of the much less atomic clustering near the surface or a better laser beam uniformity might have helped it. In the case of a longer dwell time (~$10^{-3}$ to ~$10^{-2}$ s) [37,38], the calculated Si self-diffusion length reaches ~$10^{-2}$ to ~$10^{-1}$ nm [33]. Considering the typical Si-Si bonding length in crystalline Si (~0.23 nm [39]), the Si interstitials may coalesce in this timescale. Indeed, presence and evolution of EOR defects were evidenced by weak-beam dark-field TEM [37,38].

At the end, to investigate the As activation after the 20 μs UV-LA SPR, ECVP was performed. In a typical ECVP measurement, a doped semiconductor surface is in contact with an electrolyte and forms a Schottky junction. Applying a reverse polarization field to the junction, the doped semiconductor is depleted, and a space charge zone is formed. The measured capacitance ($C$) is written as the following [22]:

$$\frac{1}{C^2} = \frac{2(V_{bi} - V)}{\varepsilon_r \varepsilon_0 q N^* A^2} \quad (1)$$

where $V_{bi}$ stands for the flat-band polarization field (called the built-in potential), $V$ for the applied polarization field, $\varepsilon_r$ and $\varepsilon_0$ for the relative and vacuum dielectric constants, $q$ for the electron charge, $N^*$ for the concentration of ionized impurities, and $A$ for the junction area. When any deep levels exist, the 1/$C^2$ versus $V$ plot shows a curvature. A fitting approach is therefore taken for all 1/$C^2$ versus $V$ curves measured at different depths after cyclic etching, involving the active shallow donor ($N_{sd}$) and inactive defect-related deep donor/acceptor ($N_{dd}$ and $N_{da}$) concentrations into $N^*$. The detailed procedure is explained elsewhere [22]. As shown in Fig. 6, the curvature observed on the 1/$C^2$ versus $V$ curve evolved with the measurement depth. A set of fitting parameters [22], which gave a good agreement with the experimental data, could be found at each depth. Fig. 7 shows the depth profile of the $N_{sd}$, $N_{dd}$, and $N_{da}$ values extracted from the fitting. The theoretical $R_{sq}$ value can be calculated from the $N_{sd}$ profile, and it is 367 ohm/sq. This value is in fact much greater than the experimentally measured one (101 ohm/sq). It implies the still high inaccuracy of our fitting parameter selection, and the real $N_{sd}$ level might become much higher (1 to 4 × $10^{20}$ cm$^{-3}$ as a rough estimation). To confirm it, a complementary analysis (e.g., differential Hall effect measurement [40]) is necessary. Anyhow, it does not affect the current discussion about the formation of deep levels after the UV-LA SPR.

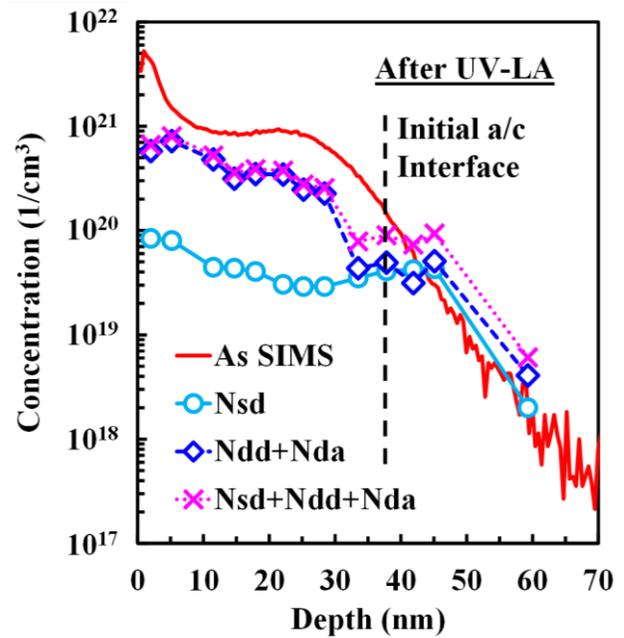

Fig. 7. Depth profiles of the shallow donor ($N_{sd}$) and deep donor/acceptor ($N_{dd}$ and $N_{da}$) concentrations extracted by the 1/$C^2$ versus $V$ curve fitting in ECVP for the 20 μs UV-LA SPR sample. The initial amorphous/crystal (a/c) interface in the SOI layer is also indicated by a dotted line.

First, it is suggested that, considering a known range of the equilibrium solid solubility limit of As in Si ($\sim 1 \times 10^{20}$ to $\sim 3 \times 10^{20}$ cm$^{-3}$ at 700 to 1100 °C [34]), many substitutional As atoms are deactivated and form deep-level defects in the regrown SOI part. As already discussed, the Si interstitials hardly travel during the UV-LA SPR. Therefore, another deactivation pathway needs to be considered. In fact, a high As dose may make it happen [41]. To begin, an As$_n$ cluster is formed among substitutional As atoms. Coupling of this cluster with a vacancy (V$_{Si}$) happens (i.e., formation of As$_n$V), emitting an interstitial Si (I$_{Si}$), especially for n = 3, 4. This emitted I$_{Si}$ replaces a nearest substitutional As atom (i.e., release of an interstitial As (I$_{As}$)). The released I$_{As}$ makes As$_n$V grow (i.e., formation of As$_{n+1}$V). This process self-continues with time and stops with As$_4$V. Indeed, As$_n$V is supposed to introduce deep donor/acceptor levels in the Si bandgap [42,43]. The kinetic Monte Carlo simulation tells that considerable As$_n$V formation may happen in the applied dwell time scale [44]. It should be noted that there is a discrepancy between the As SIMS and $N_{sd} + N_{dd} + N_{da}$ profiles. This might represent either the mean number of charges trapped by the As$_n$V defects or another As-rich defect (e.g., precipitation of As atoms although there is no direct evidence of a phase change for instance from As solid solution in the Si lattice to crystalline As).

Second, the full As activation is observed in the non-amorphized SOI part. Nevertheless, some deep donors/accepters still exist. Most of them are probably the point defects, which originally come from the ion implantation damage and are not fully annihilated during the UV-LA SPR. Indeed, V$_{Si}$ is supposed to form deep-level defects as well [42].

## IV. Conclusion

We have investigated the microsecond UV-LA induced SPR on a partially amorphized 70-nm-thick SOI substrate. The results have evidenced a single-crystal regrowth of the amorphized region by SPR at the presented cross-sectional TEM level. However, there remains some concerns to address in a future work.

First, the surface morphology was degraded during the applied UV-LA SPR. So far, two root causes are suspected: (i) a possible laser beam non-uniformity and (ii) the atomic clustering involving excess As atoms near the surface where the As segregation was observed by SIMS. Second, the excess As concentration is also supposed to trigger the deactivation of substitutional As atoms. Therefore, the initial As dose must be carefully optimized. In the case of a zero-second rapid thermal annealing followed by fast cooling (250 °C/s), the maximum carrier (i.e., substitutional As) concentration monotonically increases with increase of the annealing temperature [45]. This suggests maintaining the annealing temperature close to the upper limit (1420 K) during SPR. To further improve the As activation, introducing excess Si interstitials by pre-amorphization implantation (PAI) may work [44] because it annihilates the vacancies which are involved in the As deactivation pathway. However, as it may compete with substitutional As replacement [41], an optimal condition must be carefully examined. Fluorine co-doping may also work for gettering the vacancies [44]. Third, although the EOR defects seem hardly formed due to the negligible Si self-diffusion length, some deep levels were electrically still measured in the EOR region. This suggests suppressing the ion implantation damage to further improve the crystallinity of the EOR region after the UV-LA SPR. Cryogenic ion implantation may help it [15,46,47].

## Acknowledgment

The work covered by LASSE in this paper was supported by the IT2 project. This project has received funding from the ECSEL Joint Undertaking (JU) under grant agreement No 875999. The JU receives support from the European Union's Horizon 2020 research and innovation programme and Netherlands, Belgium, Germany, France, Austria, Hungary, United Kingdom, Romania, Israel.

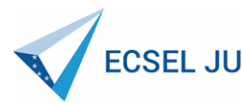
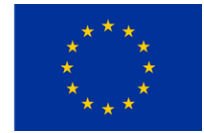